\documentclass[letterpaper,twocolumn,10pt]{article}
\usepackage{usenix2019_v3}
\usepackage{float}
\usepackage{graphicx}
\usepackage{colortbl}
\usepackage{booktabs}
\usepackage{hyperref}

\usepackage{tikz}
\def\checkmark{\tikz\fill[scale=0.4](0,.35) -- (.25,0) -- (1,.7) -- (.25,.15) -- cycle;}

\usepackage{comment}
\usepackage{balance}

\newcommand{\squishlist}{
 \begin{list}{$\bullet$}
  { \setlength{\itemsep}{0.05pt}
     \setlength{\parsep}{3pt}
     \setlength{\topsep}{3pt}
     \setlength{\partopsep}{0.1pt}
     \setlength{\leftmargin}{.9em}
     \setlength{\labelwidth}{1em}
     \setlength{\labelsep}{0.5em} } }

\newcommand{\squishend}{
  \end{list}  }

\begin{document}

\date{}

\title{\Large \bf Evaluating Snowflake as an Indistinguishable\\
  Censorship Circumvention Tool}


\author{
{\rm Kyle MacMillan}\\
Princeton University
\and
{\rm Jordan Holland}\\
Princeton University
\and
{\rm Prateek Mittal}\\
Princeton University
} 

\maketitle

\begin{abstract}
Tor is the most well-known tool for circumventing censorship. Unfortunately, Tor traffic has been shown to be detectable using deep-packet inspection. WebRTC is a popular web framework that enables browser-to-browser connections. Snowflake is a novel pluggable transport that leverages WebRTC to connect Tor clients to the Tor network.  In theory, Snowflake was created to be indistinguishable from other WebRTC services.

In this paper, we evaluate the indistinguishability of Snowflake. We collect over 6,500 DTLS handshakes from Snowflake, Facebook Messenger, Google Hangouts, and Discord WebRTC connections and show that Snowflake is identifiable among these applications with 100\% accuracy. We show that several features, including the extensions offered and the number of packets in the handshake, distinguish Snowflake among other WebRTC-based services. Finally, we suggest recommendations for improving identification resistance in Snowflake. The dataset used is publicly available. 
\end{abstract}

\section{Introduction}

Authoritarian governments continue to employ a myriad of technical mechanisms to detect and suppress internet activity~\cite{irtf-pearg-censorship-03}. Censors can trivially deny access to specific websites by blocking IP addresses or impeding DNS resolution. These techniques are only successful when the censor can accurately detect the activity it wishes to suppress.  Pluggable transports transform Tor traffic into seemingly benign traffic to disguise user activity~\cite{syverson2004tor}. We look to evaluate a new pluggable transport, Snowflake.

 \textbf{Snowflake overview.} Snowflake is composed of three core components: (1) the client, a user in a censored region, (2) the Snowflake broker, a server that connects clients to Snowflake proxies, and (3) a Snowflake proxy, a volunteer with an uncensored internet connection~\cite{fifield2017threat}. 
 After the broker has paired the client with an available proxy, the client and proxy establish a WebRTC connection. The client can then connect to a Tor relay through the Snowflake proxy.

\begin{figure}[]
\begin{center}
\includegraphics[scale=0.38]{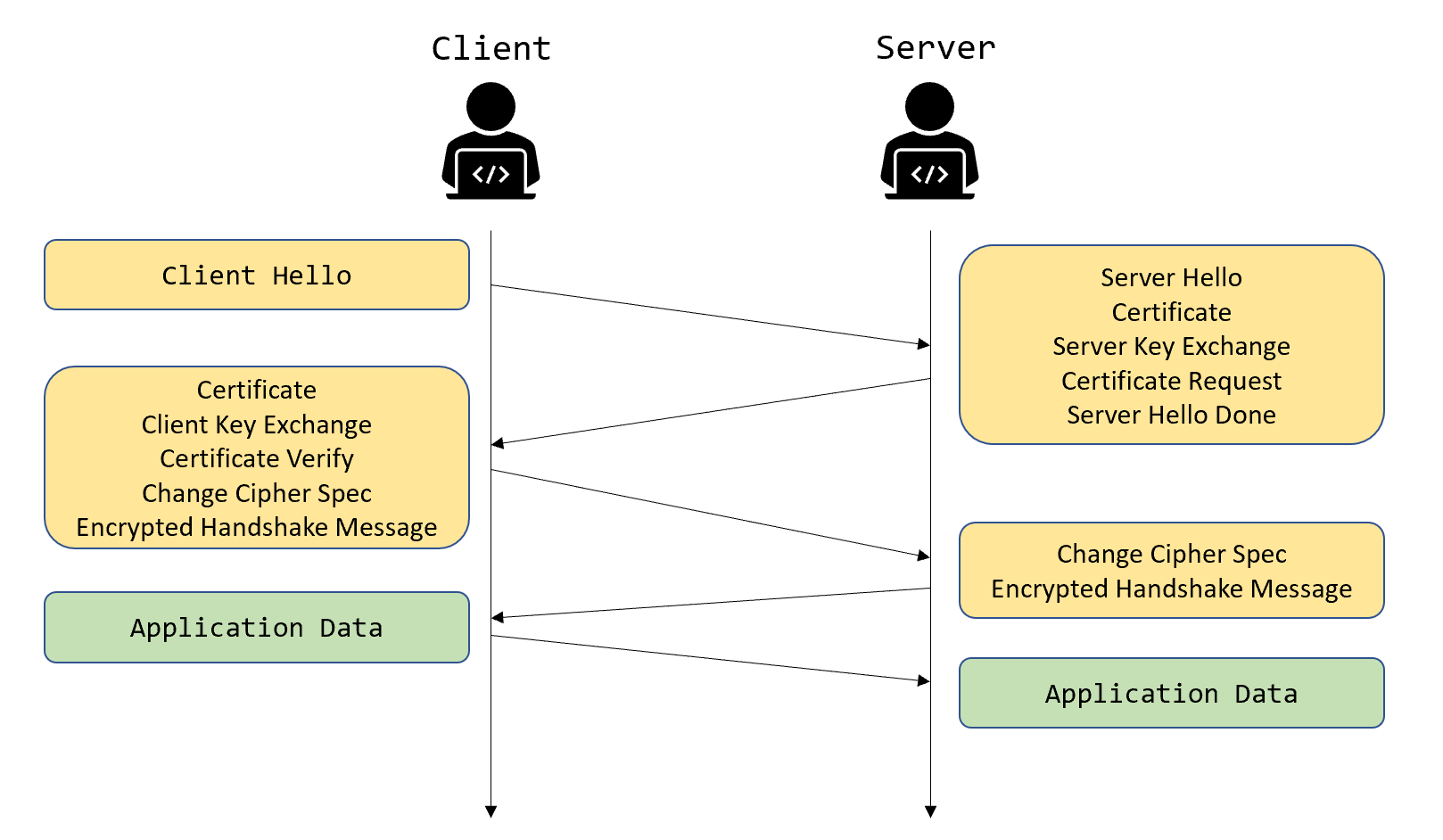}
\end{center}
\vspace{-25pt}
\caption{\label{fig:handshake} Messages exchanged during DTLS handshake}
\end{figure}

 \textbf{WebRTC connection.} WebRTC is a web framework that supports peer-to-peer communication between browsers. The WebRTC handshake utilizes either Data Transport Layer Security (DTLS) or Stream Control Transmission Protocol (SCTP). Every application examined in this work employs DTLS. As illustrated in Figure \ref{fig:handshake}, 

\textbf{Snowflake detection.} Snowflake's success relies on the ubiquity and indistinguishability of WebRTC~\cite{fifield2016fingerprintability}. If few applications use WebRTC, blocking all instances of WebRTC is a reasonable approach to blocking Snowflake. However, many web applications, such as Facebook Messenger and Google Hangouts, use WebRTC to facilitate browser to browser connections. As more services adopt WebRTC, the collateral consequences of a blanket WebRTC ban would outweigh the benefits of blocking Snowflake. In this case, one way the censor can identify Snowflake is by its WebRTC handshake~\cite{wang2015seeing}. 

\section{Evaluating Snowflake's Indistinguishability}

\textbf{Threat model.} Any surveillance state can easily observe its internet access points, with some bandwidth and computational limitations. We consider an adversary with access to a client’s WebRTC packets, including headers, protocols, and payloads. We assume that Snowflake is resistant to detection by IP address. 
Given a large volume of temporary proxies and the use of techniques such as domain fronting, a censor's ability to detect Snowflake connections via IP addresses is limited~\cite{fifield2015blocking}.

\textbf{Data collection.} We collect data by capturing isolated DTLS handshakes. Table \ref{table:data outline} summarizes the handshakes collected. We have released this dataset so that it is publicly available.~\cite{macmillan_snowflake_data}
\begin{table}[H]
\begin{tabular}{lcccc}
\hline
               & \multicolumn{1}{l}{\textbf{Snowflake}} & \multicolumn{1}{l}{\textbf{Facebook}} & \multicolumn{1}{l}{\textbf{Google}} & \multicolumn{1}{l}{\textbf{Discord}} \\ \hline
\rowcolor[HTML]{E2E2E2} 
Firefox        & 991     & 796     & 1000       & 992                                  \\
Chrome         & 0   & 784    & 995         & 997                                                                    \\
\midrule
\textbf{Total} & 991      & 1580        & 1995           & 1989                                 
\end{tabular}
\caption{Number of handshakes collected for each services on a given browser.}
\label{table:data outline}
\end{table}

\textbf{Average Packets per Handshake.} Immediately observable is the difference between the average number of packets sent per handshake among the services, as shown in Table \ref{table:avg packets}. The Snowflake handshake tends to require several retransmissions, resulting in a much longer handshake than other services, where retransmissions are observed sparingly. 

\begin{table}[H]
\begin{tabular}{cccc}
\hline
                \multicolumn{1}{l}{\textbf{Snowflake}} & \multicolumn{1}{l}{\textbf{Facebook}} & \multicolumn{1}{l}{\textbf{Google Hangouts}} & \multicolumn{1}{l}{\textbf{Discord}} \\ \hline
\rowcolor[HTML]{E2E2E2} 
         13.42     & 4.4     & 4.5       & 5.6                                                                                       \\
\midrule
                              
\end{tabular}
\caption{Average number of handshakes collected for each service on a given browser.}
\label{table:avg packets}
\end{table}

\section{Classifying Handshakes}
\textbf{Classification Methods} Table \ref{table:features} summarizes the 20 features extracted. We use one-hot-encoding to transform non-numeric data into binary features. We choose a random forest classifier because it allows us to examine which features drive model performance. We use 5-fold cross validation for all evaluation metrics. We evaluate our classifier using accuracy and micro-weighted F1 scores.

\begin{table}[H]
\begin{tabular}{lcc}
\hline
Feature             & \multicolumn{1}{l}{\textbf{Client Hello}} & \multicolumn{1}{l}{\textbf{Server Hello}} \\ \hline
\rowcolor[HTML]{E2E2E2} 
Length              & \checkmark                                         & \checkmark                                         \\
Message Sequence    & \checkmark                                         & \checkmark                                         \\
\rowcolor[HTML]{E2E2E2} 
Fragment Offset     & \checkmark                                         & \checkmark                                         \\
DTLS Version        & \checkmark                                         & \checkmark                                         \\
\rowcolor[HTML]{E2E2E2} 
SID Length          & \checkmark                                         & \checkmark                                         \\
Cookie Length       & \checkmark                                         &                                           \\
\rowcolor[HTML]{E2E2E2} 
Cipher Suite Length & \checkmark                                         &                                           \\
Cipher Suites       & \checkmark                                         &                                           \\
\rowcolor[HTML]{E2E2E2} 
Extension Length    & \checkmark                                         & \checkmark                                         \\
Extension           & \checkmark                                         & \checkmark                                         \\
\rowcolor[HTML]{E2E2E2} 
Cipher Chosen       &                                           & \checkmark                                        
\end{tabular}
\caption{Features extracted from WebRTC handshakes.}
\label{table:features}
\end{table}

\textbf{Classification evaluation.}  One-hot-encoding the features in Table \ref{table:features} produces 61 total features.
We train a classifier using scikit learn's Random Forest Classifier module~\cite{barradas2018effective}. Across all classes the average accuracy is 99.8\% and the classifier perfectly identifies Snowflake in terms of accuracy and micro-weighted f1 score.  Given these results, we search for \emph{identifiers}: features whose values are unique to each class. 

\textbf{Analyzing feature importance.} We leverage the model's feature importances to search for Snowflake identifiers. Table \ref{table:identifiers} shows features unique to Snowflake. \emph{supported\_groups} and \emph{renegotiation\_info} are extensions offered in the Server Hello. \emph{Server Message Sequence: "1"} indicates that the Snowflake DTLS protocol includes optional \emph{Client Hello} and \emph{Hello Verify Request} packets that the other services omit. 

\begin{table}[H]
\begin{center}
\begin{tabular}{lrrrr}
\hline
\toprule
\multicolumn{4}{r}{Application} \\
 \cmidrule(r){2-5}
                 Feature     & \textbf{SF} & \textbf{FB} & \textbf{G} & \textbf{D} \\ \midrule
\rowcolor[HTML]{E2E2E2} 
Server Message Sequence: 1           & 100         & 0           & 0          & 0          \\
renegotiation\_info          & 0           & 100         & 100        & 100        \\
\rowcolor[HTML]{E2E2E2} 
supported\_groups & 100           & 0          & 0        & 0               
\end{tabular}
\caption{Percentage of handshakes that contained a given feature for Snowflake (SF), Facebook Messenger (FB), Google Hangouts (G), and Discord (D).}
\label{table:identifiers}
\end{center}
\end{table}


\section{Recommendations}

Based on our results, it is necessary to modify the Snowflake WebRTC implementation to resist detection by content. The following modifications are short-term fixes that can improve Snowflake's indistingiushability:
\squishlist
    \item Do not send the optional \emph{Client Hello} and \emph{Hello Verify Request} from the DTLS handshake
    \item Offer ‘renegotiation\_info’ as an extension in the server hello
    \item Do not offer ‘supported\_groups’ as an extension in the server hello
\squishend

However, modifying Snowflake's WebRTC approach to mimic popular services may be futile~\cite{houmansadr2013parrot}. As a long-term solution, we suggest that Snowflake use an existing WebRTC-based service's implementation~\cite{houmansadr2013want}.

\bibliographystyle{unsrt}
\bibliography{refs.bib}
\balance

\end{document}